\documentclass[prd,aps,showpacs,tightenlines,superscriptaddress,amsmath,amssymb,amsfonts,twocolumn]{revtex4}
\usepackage{amssymb,amsmath,amsthm,amsbsy,epsfig,color,graphicx,times}
\usepackage[ansinew]{inputenc}
\usepackage[english]{babel}
\usepackage{color}
\usepackage{braket}
\usepackage{booktabs}
\usepackage{tabularx}
\usepackage{array}

\begin{document}

\title{Probing axion mediated fermion--fermion interaction by means of entanglement}

\author{A. Capolupo}
\email{capolupo@sa.infn.it}
\affiliation{Dipartimento di Fisica ``E.R. Caianiello'' Universit\`{a} di Salerno, and INFN -- Gruppo Collegato di Salerno, Via Giovanni Paolo II, 132, 84084 Fisciano (SA), Italy}

\author{G. Lambiase}
\email{lambiase@sa.infn.it}
\affiliation{Dipartimento di Fisica ``E.R. Caianiello'' Universit\`{a} di Salerno, and INFN -- Gruppo Collegato di Salerno, Via Giovanni Paolo II, 132, 84084 Fisciano (SA), Italy}

\author{A. Quaranta}
\email{anquaranta@unisa.it}
\affiliation{Dipartimento di Fisica ``E.R. Caianiello'' Universit\`{a} di Salerno, and INFN -- Gruppo Collegato di Salerno, Via Giovanni Paolo II, 132, 84084 Fisciano (SA), Italy}

\author{S. M. Giampaolo}
\email{sgiampa@irb.hr}
\affiliation{Institut Ru\dj er Bo\v{s}kovi\'c, Bijeni\v{c}ka cesta 54, 10000 Zagreb, Croatia}

\begin{abstract}

We propose a new approach in the investigation and detection of axion and axion--like particles based on the study of the entanglement for two interacting fermions.
We study a system made of two identical fermions with \mbox{spin $-1/2$}, and we show that fermion--fermion interaction mediated by axions leads to a non--zero entanglement between the fermions. An entanglement measurement can reveal the interaction, providing an indirect evidence of the existence of axions. We discuss how the other interactions affect the entanglement, and how to isolate the axion contribution. Particular care is devoted to the analysis of the magnetic dipole--dipole interaction, which turns out to be,  apart from axions, the most relevant contribution to the entanglement, and we show that it can be suppressed by setting opportunely the duration of the observation. We also introduce a two--body correlation function, which could be directly observed in an experiment, and plays the role of an entanglement witness.

\end{abstract}

\maketitle

\section{Introduction}

The standard model provides a simple explanation for a wide range of phenomena involving fundamental particles and their interactions.
In spite of its success, it does not represent the definitive theory of elementary particles. Several phenomena, ranging from particle mixing~\cite{intro1,intro2,intro3,intro4,intro5} to the quantum features of gravitation~\cite{Ellis2009}, are beyond the standard model.
Among the shortcomings of the theory is the so--called \emph{strong CP problem} in Quantum Chromodynamics (QCD)~\cite{PDG2019,Peccei1977,Peccei2008}.
The QCD Lagrangian features a gluon--gluon interaction term that in principle allows for an arbitrary violation of CP symmetry, whereas no such violation is observed in strong processes~\cite{PDG2019,Peccei1977,Peccei2008,Wilczek1978,Weinberg1978}.
As a direct consequence of the CP violating term, one would expect a relatively large neutron electric dipole moment, which the recent experiments constrain below $3 \times 10^{-26} e \, cm $  (see~\cite{Pendlebury2015} and, for a more recent result~\cite{Abel2020}).  Similar bounds on the electric dipole moment of atoms and molecules ~\cite{Graner2016,Flambaum2019} pose a further constraint on the magnitude of the CP violating term.
To remedy this inconsistency, R. Peccei and H. Quinn introduced a new global symmetry $U_{PQ}(1)$ (called Peccei--Quinn symmetry) that is spontaneously broken~\cite{Peccei1977,Peccei2008}.
As shown by Frank Wilczek and Steven Weinberg~\cite{Wilczek1978,Weinberg1978}, this results in a new particle, named the axion, which is the pseudo--Nambu--Goldstone boson of the broken $U_{PQ} (1)$ symmetry~\cite{Raffelt2007}.

The scale at which the symmetry breaking occurs $f_A$, known as the axion decay constant, determines, according to the model considered, both the axion mass $m_a \propto \frac{1}{f_A}$ and the effective couplings with the standard matter.
The original proposal considered $f_A$ close to the electroweak scale~\cite{Peccei1977,Wilczek1978,Weinberg1978}, a hypotesis that was later ruled out by the experiments.
Soon alternative axion models were devised, notably the KSVZ model~\cite{KSVZ}, featuring heavy quarks carrying a Peccei--Quinn (PQ) charge, and the DFSZ model~\cite{DSFZ}, in which the ordinary quarks and additional Higgs doublets carry the PQ charge.
They provide a reference for two large classes of axion models (\emph{hadronic} and \emph{GUT} axions).
Today the axion decay costant is estimated to be very large $f_A > 10^{9} GeV$, so that the QCD axions (both hadronic and GUT) have to be very light $m_a \sim 10^{-6} - 10^{-2} eV$ and very weakly interacting~\cite{PDG2019}.
These aspects make axions a natural candidate for dark matter.

Moreover, motivated by the search for dark matter components, a variety of Axion--like--particles (ALPs) has been introduced.
They can deviate significantly from the original Peccei--Quinn proposal, and are not necessarily tied to the solution of the strong CP problem, but share the nature of axions and are weakly interacting with the standard matter.
In these models the relation between the coupling constant and the mass of the ALP can differ from the direct proportionality that characterizes the PQ axions.
They range from masses $m_{a}\sim 10^{-22}$ eV, characteristic of the ultra--light axions~\cite{Kim2016,Demartino2017}, up to masses of $1$ TeV for the heavy GUT axions~\cite{Rubakov1997}. Axions and ALPs are, to date, one of the most credible explanations for dark matter~\cite{Marsch2016,Lambiase2018,Auriol2019,Houston2018}.

Driven by large theoretical interest, several experiments were designed to prove the existence of ALPs. Perhaps unsurprisingly, considered their extremely small interaction rates, the experimental search for ALPs has proven to be very challenging.
Many experiments take advantage of the axion--photon coupling.
Among them, searches for polarization anomalies in the light propagating through a magnetic field (PVLAS)~\cite{PVLAS}, ``light shining through a wall'' experiments (OSQAR, ALPS)~\cite{OSQAR,OSQAR2,ALPS}, detection via the Primakoff effect (CAST)~\cite{CAST}, and haloscope experiments (ADMX)~\cite{ADMX}.
More recently other approaches based on geometric phases~\cite{Capolupo2015} and QFT effects in the axion--photon mixing~\cite{Capolupo2019} were also suggested.
Astrophysical observations and terrestrial experiments over the decades have restricted the allowed regions in parameter space, and further constraints might come from the analysis of the axion--nucleon and axion--lepton interactions, as suggested for instance in~\cite{Bezerra2016}. Experiments based on the axion--fermion interaction have also been proposed (QUAX)~\cite{QUAX}.
Despite this, no evidence for the existence of the ALPs has been found up to now.

In this paper, we propose a different approach to the detection of axions, based on the study of the entanglement arising between two fermions interacting via axion exchange.
In recent years, theoretical and experimental analysis of the entanglement properties have found application in the most disparate fields, from quantum biology~\cite{Arndt2009} to statistical physics~\cite{Vidal2003,Amico2008}, and also as a tool to gain insight on fundamental interactions, like gravity~\cite{Bose2017,Marletto2017,Simonov2019,Marshman2019}.
These applications stem from the fact that the emergence of entanglement between two (or more) physical objects is strictly connected to the presence of a quantum interaction between them acting as a quantum channel~\cite{Giampaolo2019,Bennett1999}.
Consequently, an analysis of the entanglement properties provides information about the interaction itself.

We focus on the axion--mediated fermion--fermion interaction, which assumes the familiar form of a Yukawa coupling between the pseudoscalar field and two fermionic fields.
In the non--relativistic limit, the axion--induced interaction reduces to an effective two--body potential~\cite{Moody1984,Daido2017}, that acts as a source of entanglement for the two fermions.
An entanglement measurement might then provide evidence for the pseudoscalar interaction, and thus for the existence of axions/ALPs.
Obviously, the two fermions interact with each other in many other ways, i.e. gravitationally, magnetically, etc.
All these interactions are potentially a source of entanglement.
Hence, one of the main goals of this work is to show how it is possible to extract the axions--induced entanglement contribution from the others.
We choose to quantify the entanglement through the 2--Renyi entropy~\cite{Horodecki2002,Plenio2007,Nielsen2011,Giampaolo2013} that has the advantage to be, in several systems, directly connected to experimentally accessible quantities~\cite{Bovino2006,Brydges2019,Lesche04,Islam15,Abanin12}.
Since a direct measurement of the entanglement entropy is challenging in many cases, here we individuate a two--body spin correlation function that plays the role of an entanglement witness, and can be more easily accessed.

An entanglement witness is a physical quantity that is strictly related to the family of states under analysis, with the property of vanishing simultaneously with the entanglement.
The detection of a non--zero value of the entanglement witness therefore implies that the entanglement is non--zero.
Specialized to our case, under suitable conditions, a non--vanishing witness would signal the presence of axions, and then provide an indirect evidence for their existence.
For masses in the range of $(10^{-3} - 1)$ eV and coupling constants close to the actual constraints, that are compatible with some ALPs models, the witness is significantly different from zero.

The paper is organized as follows.
We first recap the axion--fermion pseudoscalar interaction and the corresponding two--fermion potential in the non--relativistic limit (Sec II).
From the knowledge of the potential we compute the time--dependent entanglement between the two fermions, and we individuate a specific time at which, in absence of axions, the entanglement must vanish.
If in correspondence with such time the entanglement is different from zero, the presence of axions is detected (Sec III).
Soon after we introduce an entanglement witness and we present a numerical analysis (Sec IV), and finally we draw our conclusions
(Sec V).

\section{Fermion--Fermion interaction induced by Axions}

Let us start by recalling the main features of the axion--mediated fermion--fermion interaction.
The coupling of axions with fermions is described by a Yukawa pseudoscalar vertex~\cite{Moody1984,Daido2017}.
If $\phi$ is the axion field and $\psi_1,\psi_2$ are the fermion fields, the interaction term reads
\begin{equation}\label{YukawaInteraction}
  \mathcal{L}_{INT} = -  \sum_{j=1,2}i g_{pj} \phi \bar{\psi}_{j}  \gamma_{5} \psi_{j}
\end{equation}
where $\gamma_{5}$ is the product of Dirac matrices $i\gamma^{0} \gamma^{1} \gamma^{2} \gamma^{3}$ and $g_{pj}$ are the effective axion--fermion coupling constants, which depend critically on the fermions considered and the underlying axion (or ALP) model.
Since the couplings are expected to be small, i.e. \mbox{$g_{pj} \ll 1$}, the scattering amplitudes can be well approximated by the leading order in the perturbative expansion.
For the scattering $\psi_1(p_1)\psi_2(p_2) \rightarrow \psi_1(p^{'}_1) \psi_2(p^{'}_2)$ we have
\begin{equation}
 \label{YukawaInteraction1}
 \imath \mathcal{A}\! =\! \bar{u}_1^{s'_1}(p^{'}_1) g_{p1} \gamma_5 u_1^{s_1}(p_1) \frac{i}{q^2 \! - \! m^2}\bar{u}_1^{s'_2}(p^{'}_2) g_{p2} \gamma_5 u_2^{s_2}(p_2) \!\!
\end{equation}
where the pseudoscalar free propagator with momentum \mbox{$q = p_1^{'}-p_1=p_2-p_2^{'}$} appears, and  $m$ is the axion mass.
Here $u_i^{s_i}(p_i)$ are the solutions of the free Dirac equation in momentum space, for the i-th fermion with momentum $p_i$ and spin projection $s_i$:
\begin{equation}
u^{s_i}_i (p_i) = \sqrt{\frac{\omega_i + M_i}{2 \omega_i}} \begin{pmatrix} \phi_{s_i} \\ \frac{\pmb{\sigma} \cdot \   \pmb{p}_i}{\omega_i + M_i} \phi_{s_i} \end{pmatrix}
\end{equation}
with $M_i$ mass of the $i$-th fermion, $\omega_i = \sqrt{\pmb{p}_i^2 + M_i^2}$ energy of the $i$-th fermion and $\phi_{s_i}$ normalized two--component spinors. In the non--relativistic limit $\omega_i \approx M_i$, these become
\begin{equation}\label{NonrelativisticSpinors}
u^{s_i}_i (p_i) \approx  \begin{pmatrix} \phi_{s_i} \\ \frac{\pmb{\sigma} \cdot \   \pmb{p}_i}{2 M_i} \phi_{s_i} \end{pmatrix} \ .
\end{equation}
Inserting equations \eqref{NonrelativisticSpinors} in equation \eqref{YukawaInteraction1}, one obtains the scattering amplitude for non--relativistic fermions
\begin{equation}\label{NonrelativisticAmplitude}
\mathcal{A} \approx \frac{g_{p_1}g_{p_2}}{\pmb{q}^2 + m^2} \frac{\left(\phi_{s'_1}^{\dagger} \left(\pmb{\sigma} \cdot \pmb{q}\right) \phi_{s_1} \right) \left( \phi_{s'_2}^{\dagger} \left(\pmb{\sigma} \cdot \pmb{q}\right) \phi_{s_2} \right)}{4M_1M_2} \ .
\end{equation}
By Fourier transforming the amplitude \eqref{NonrelativisticAmplitude} into real space, one finds the two--body potential due to axion exchange ~\cite{Daido2017}
 \begin{eqnarray}
 \label{Pseudoscalar_Potential}
 V (\pmb{r}) \!&\!=\!&\!-\frac{g_{p_1} g_{p_2} e^{-m r}}{16 \pi M_1 M_2}\left[ {\pmb{\sigma}}_1 \cdot {\pmb{\sigma}}_2
\left(\frac{m}{r^2}+\frac{1}{r^3}+\frac{4}{3}\pi \delta^3({\bf{r}})\right)\right. \nonumber \\
\!&\! \!&\;\;\;\;\;\;\;\;\; \left. - \left({\pmb{\sigma}}_1  \cdot \pmb{\hat{r}}\right)
\left({\pmb{\sigma}}_2 \cdot \pmb{\hat{r}}\right)
\left(\frac{m^2}{r}+\frac{3m}{r^2}+\frac{3}{r^3}\right)\right] \ ,
\end{eqnarray}
where $r$ ($\pmb{\hat{r}}$) stands for the modulus (the unit vector) of the relative distance between the fermions and $\delta^3({\bf{r}})$ is the Dirac delta, while $\pmb{\sigma}_i$ is the three--dimensional vector of Pauli operators defined on the $i$--th fermion. For two identical non--relativistic fermions one has $M_1 = M_2 = M$ and $g_{p_1} = g_{p_2} = g_{p}$, yielding the interaction Hamiltonian
\begin{eqnarray}
 \label{Pseudoscalar_Hamiltonian}
 H_p\!&\!=\!&\!- \frac{g_p^2 e^{-m r}}{16 \pi M^2}\left[ {\pmb{\sigma}}_1 \cdot {\pmb{\sigma}}_2
\left(\frac{m}{r^2}+\frac{1}{r^3}+\frac{4}{3}\pi \delta^3({\bf{r}})\right)\right. \nonumber \\
\!&\! \!&\;\;\;\;\;\;\;\;\; \left. - \left({\pmb{\sigma}}_1  \cdot \pmb{\hat{r}}\right)
\left({\pmb{\sigma}}_2 \cdot \pmb{\hat{r}}\right)
\left(\frac{m^2}{r}+\frac{3m}{r^2}+\frac{3}{r^3}\right)\right].
\end{eqnarray}
In the following we will always consider $r$ large enough that the contact term proportional to $\delta^3(\pmb{r})$ can be neglected. Assuming, in addition, that $\pmb{\hat{r}}$ coincides with the $z$--direction, we obtain
\begin{eqnarray}
 \label{Pseudoscalar_Hamiltonian_2}
 H_p\!&\!= - \!&\!\frac{g_p^2  e^{-m r}}{16 \pi M^2r^3 }\left[
  m^2 r^2\, \sigma^z_1\sigma^z_2 +\left(mr+1\right)\mathcal{O}
\right]
\end{eqnarray}
where the operator $\mathcal{O}$  is defined as $\mathcal{O}=2 \sigma^z_1\sigma^z_2 -\sigma^x_1\sigma^x_2 -\sigma^y_1\sigma^y_2$. The equation \eqref{Pseudoscalar_Hamiltonian_2} is the interaction, due to axion exchange, between two identical non--relativistic fermions, and represents the starting point for our analysis.

\section{Dynamics of entanglement}

Since we wish to analyze the entanglement between two fermions due to the axion--mediated interaction \eqref{Pseudoscalar_Hamiltonian_2}, we first need to establish whether the latter can induce entanglement in a fermionic system. More precisely, considering two fermions interacting through the Hamiltonian $H_p$ of eq. \eqref{Pseudoscalar_Hamiltonian_2} and initially prepared in a fully separable (non--entangled) state, we need to determine whether their state develops a non-vanishing entanglement under the action of $H_p$.
There exist precise conditions that a Hamiltonian has to fulfill in order to produce entanglement: 1) a non completely degenerate spectrum; 2) the impossibility to be reduced to the sum of local terms acting separately on every single object~\cite{Giampaolo2019,Simonov2019,Bennett1999}.
As it is easy to check, the Hamiltonian in eq.~\eqref{Pseudoscalar_Hamiltonian_2} fulfills both the requirements, and shall induce entanglement on the two fermion system. This fact can be used to reveal axions and ALPs through the analysis of the entanglement properties of the two fermion state.

Therefore we focus on a system of two identical spin--$\frac{1}{2}$ fermions (for instance electrons or neutrons) and study the time evolution of its entanglement properties. We write the state of the system as
\begin{equation}\label{Fullstate}
  \Psi (\pmb{r}_1, \pmb{r}_2, s_1, s_2; t) = R(\pmb{R}, \pmb{r}; t) \psi(s_1,s_2;t)
\end{equation}
where the spatial wave--function $R$ depends on the center of mass position $\pmb{R}$ and the relative position of the two fermions $\pmb{r} = \pmb{r}_1 - \pmb{r}_2$, while the spin wave--function $\psi(s_1,s_2;t)$ is a state vector in the product space $H^{spin}_1 \otimes H^{spin}_2$ of the spin Hilbert spaces associated to the two particles. In order to simplify the analysis, we assume, as a first approximation, that the spatial wave--function $R$ is sharply peaked at a given value of the distance $r = |\pmb{r}_1 - \pmb{r}_2|$, and so remains during the time interval of interest. We then consider the distance $r$ in eq. \eqref{Pseudoscalar_Hamiltonian_2} as a parameter, and the Hamiltonian $H_p$ as an operator acting on the spin state alone. Of course, the full wave--function $\Psi$ must be antisymmetric under particle exchange. Since we shall consider symmetric spin states $\psi(s_1,s_2;t)$, the spatial wave--function $ R(\pmb{R}, \pmb{r}; t)$ must be antysimmetric.

At $t=0$ we assume that the spin state of the whole system is fully separable, i.e. that it can be written as the tensor product of two states each defined on a single fermion.
In other words we have that, at $t=0$, the state of the system can be written as
\begin{eqnarray}
 \label{initial_state}
 \ket{\psi (0)}&=&\ket{\varphi(0)}_1 \otimes \ket{\varphi(0)}_2 \\
 \ket{\varphi(0)}_i &=&  \cos(\theta) \ket{\uparrow}_i + e^{\imath \phi } \sin(\theta) \ket{\downarrow}_i\nonumber
\end{eqnarray}
%
where we have dropped the spin arguments $s_1, s_2$ and have switched to a more convenient Dirac notation. Here $\ket{\uparrow}_{i}$ and $\ket{\downarrow}_{i}$ denote the eigenstates of the magnetic moment along the direction joining the two fermions \mbox{($z$-direction)}. Being the state of Eq. \eqref{initial_state} separable, the entanglement vanishes for \mbox{$t=0$}.

If the two fermions would interact with each other only through axions, a non--vanishing entanglement would directly signal their presence, but this is not the case.
The two fermions generally interact with each other through several channels, and any of these is potentially a source of entanglement, depending on the distance $r$ and on the states that one considers.

As we are interested in highlighting the entanglement due to axions, these interactions produce an unwanted contribution that has to be minimized in order to successfully reveal the axion--induced entanglement. Then we must devise a setting in which these additional sources of entanglement are suppressed or significantly reduced.

At first, let us deal with the weak and the strong nuclear interactions.
They are relevant only for very small distances and, hence, assuming, in our setup, a relative distance large enough ($r > 10^{-12} m$), we can neglect them altogether.
Secondly, let us consider the gravitational and the electrostatic (if the fermions have a non--vanishing electric charge) interactions.
Their action on the evolution of the spin state cannot induce entanglement and amounts to a global phase factor.
Indeed, as we have previously said, a fundamental requirement for the interaction to induce entanglement in a system, is that the spectrum of the associated Hamiltonian is not fully degenerate.
Particles such as electrons, protons, or neutrons are characterized by precise values of charge and mass that do not depend on their spin states.
Therefore, any state depending solely on the spin of the particles, like the one we are considering, will react to gravitational and electrostatic interactions in the same way and to not generate entanglement.

Another unwanted entanglement source is
 the dipole--dipole magnetic interaction, whose Hamiltonian reads
\begin{eqnarray}
 \label{Magnetic_Hamiltonian}
 H_\mu=-\frac{1}{4\pi r^3}\frac{g^2 {q_e}^2}{16 \, M^2} \mathcal{O}\,,
\end{eqnarray}
where $g$ is the g--factor and $q_e$ is the charge of the electron. As done in Eq. \eqref{Pseudoscalar_Hamiltonian_2}, we have dropped a contact term proportional to $\delta^{3} (\pmb{r})$ in Eq. \eqref{Magnetic_Hamiltonian}, assuming that $r$ is large enough.
Being associated with the exchange of massless photons, this interaction is not confined at short--range and it is sensible to the different spin states in a way that cannot be reduced to the action of local operators.
Hence its contribution to the entanglement is different from zero.
However, in the following we will show that,  by properly setting the time interval of the entanglement measurement,
 the contribution to entanglement due to  the dipole--dipole magnetic interaction can be neglected and  the entanglement
reduces to the axion contribution alone.

Starting from the initial state $\ket{\psi (0)}$, the state at $t>0$ can be obtained as $\ket{\psi(t)}=U(t)\ket{\psi(0)}$, where the time evolution operator $U(t)$ is unitary since we consider our system to be closed (we neglect any other interaction with the surrounding world).
$U(t)$ can be written as $U(t)=\exp[-\imath t (H_T)]$, where the total Hamiltonian $H_T$ is the sum of the magnenetic $(H_\mu)$ and the axion term $(H_p)$, i.e. $H_T=H_p+H_\mu$, and reads
\begin{eqnarray}
 \label{Total_Hamiltonian}
 \!H_T\!&\!=\!&\!-\frac{A}{r^3} \left[
 \mathcal{O} + B e^{-m r} \left(m^2r^2\, \sigma^z_1\sigma^z_2  +\mathcal{O} \left(mr+1\right)\right)
 \right].\,\,\,\,
\end{eqnarray}
In eq.~(\ref{Total_Hamiltonian}) the parameter $A=\frac{g^2 {q_e}^2}{64\pi M^2}$ is the strength of the magnetic interaction while $B=\frac{4 g_p^2 }{g^2 q_e^2}=\frac{4 g_p^2 }{\alpha g^2}$ quantifies the relative weight of the axion interaction and $\alpha$ denotes the fine structure constant.

Since the operator $U(t)$ is unitary for any time $t\ge 0$, the state $\ket{\psi(t)}$ remains a pure state, although in general, differently from $\ket{\psi(0)}$, it is entangled.
The amount of entanglement between the two fermions in $\ket{\psi(t)}$ can be quantified using different measures. In the present work we consider the $2$--Renyi entropy~\cite{Nielsen2011,Horodecki2002,Plenio2007,Giampaolo2013}, that has the advantage, with respect to the other entropy--based entanglement measures to be associated, at least in some experimental devices, to experimentally accessible quantities~\cite{Lesche04,Islam15,Abanin12,Bovino2006,Brydges2019}. The \mbox{$2$--Renyi} entropy is defined as $S_2=-\ln(\mathcal{P}(\rho_i(t)))$, where \mbox{$\rho_i(t)=\mathrm{Tr_{j \neq i}}(\ket{\psi(t)}\bra{\psi(t)})$} is the reduced density matrix obtained projecting $\ket{\psi(t)}$ on the Hilbert space defined on one of the two femions and \mbox{$\mathcal{P}(\rho_i(t))=\mathrm{Tr} \rho^2_i(t)$} is the purity of $\rho_i(t)$.
In our system the $2$-Renyi entropy reads
\begin{eqnarray}
 \label{2Renyi_entropy}
S_2(t)\!&\!=\!&\!-\!\ln \left[\!1\!-\frac{\sin^4(2\theta)}{2} \sin^2(\Gamma t)\right]
\end{eqnarray}
where $\Gamma = \frac{6 A}{r^3}\left[1+\frac{B}{3} e^{-r m}\left(3+3 rm+r^2m^2\right)\right]$.
As we can see from the expression of $\Gamma$, the entanglement derives from both the dipole--dipole magnetic interaction
(due to  $\frac{6 A}{r^3}$)  and from the presence of the axions (due to the term $\frac{2 A}{r^3}\left[ B e^{-r m}\left(3+3 rm+r^2m^2\right)\right]$ ). Given the form of the \mbox{$2$--Renyi} entropy \eqref{2Renyi_entropy}, there exist certain times at which the entanglement is only due to the axion--mediated interaction. Indeed, by setting
$t=n t^*$, with $n$ a positive integer and \mbox{$ t^*=\frac{\pi r^3}{6 A}$}, we obtain:
\begin{eqnarray}\label{2Renyi_entropy_tstar}
S_2 (n t^*) &=& -\ln \Bigg{\{} 1 - \frac{\sin^4 (2\theta)}{2} \times  \\
&\times & \sin^{2} \left[ n\pi \left( 1 + \frac{B}{3} e^{- mr} (3 + 3mr + m^2r^2) \right) \right] \Bigg{\}} \nonumber \, \\
&\simeq& \frac{\sin^4 (2\theta)}{2} n^2 \pi^2 B^2  e^{-2mr}\! \left(1 + mr + \frac{m^2r^2}{3}\right)^2 . \nonumber
\end{eqnarray}
In equation \eqref{2Renyi_entropy_tstar}, the dependence on the dipole--dipole magnetic interaction has disappeared. Recalling that $B \propto g_p^2$, we can see at once that if there is no axion--mediated interaction ($g_p \rightarrow 0$), the $2$-Renyi entropy, and thus the entanglement, vanishes.
More precisely, as $g_p \ll 1$, the last line of eq.~\eqref{2Renyi_entropy_tstar} shows that the entropy is proportional to $B^2$ and then to $g_p^4$. In the case in which a non--zero entanglement is detected, in correspondence with the times $t = n t^*$, one can conclude that the former is a consequence of the axion--induced interaction alone. This would constitute an indirect proof of the existence of axions.

\begin{figure}
\centering
\includegraphics[width=.95\linewidth]{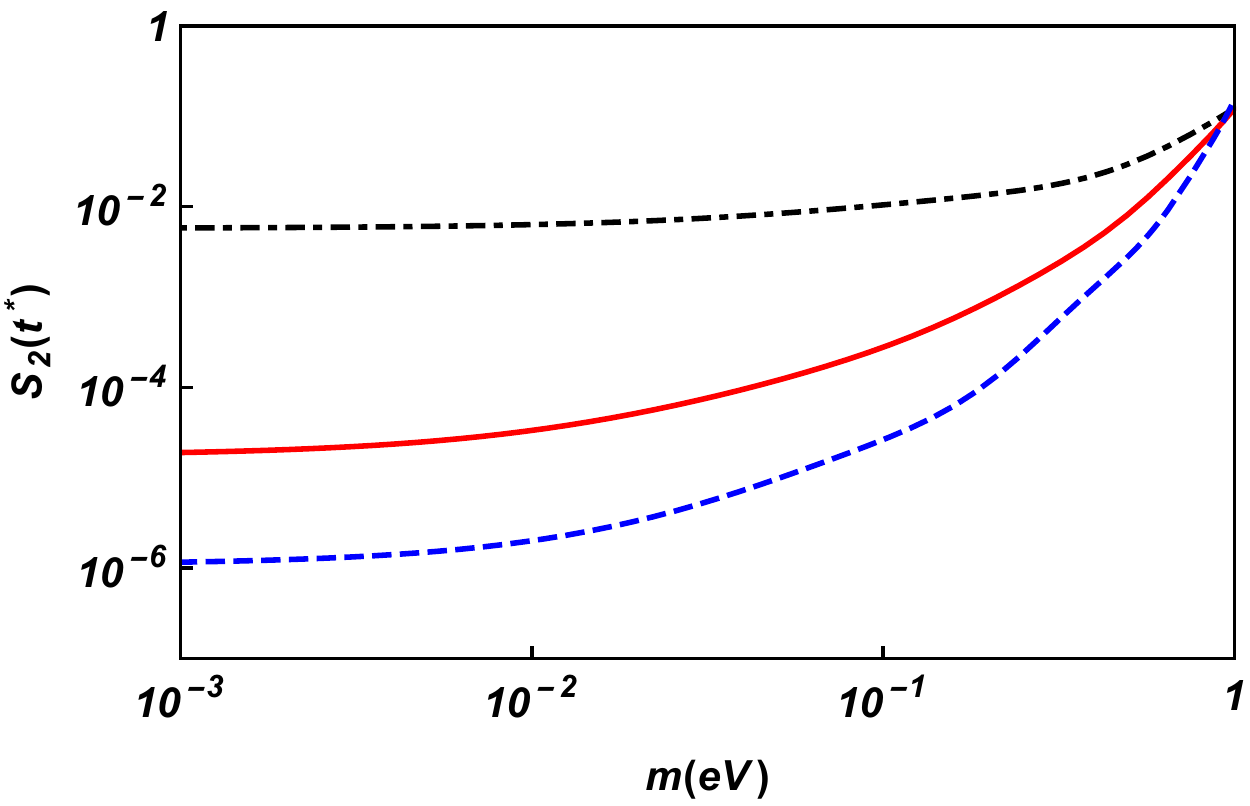}
\includegraphics[width=.95\linewidth]{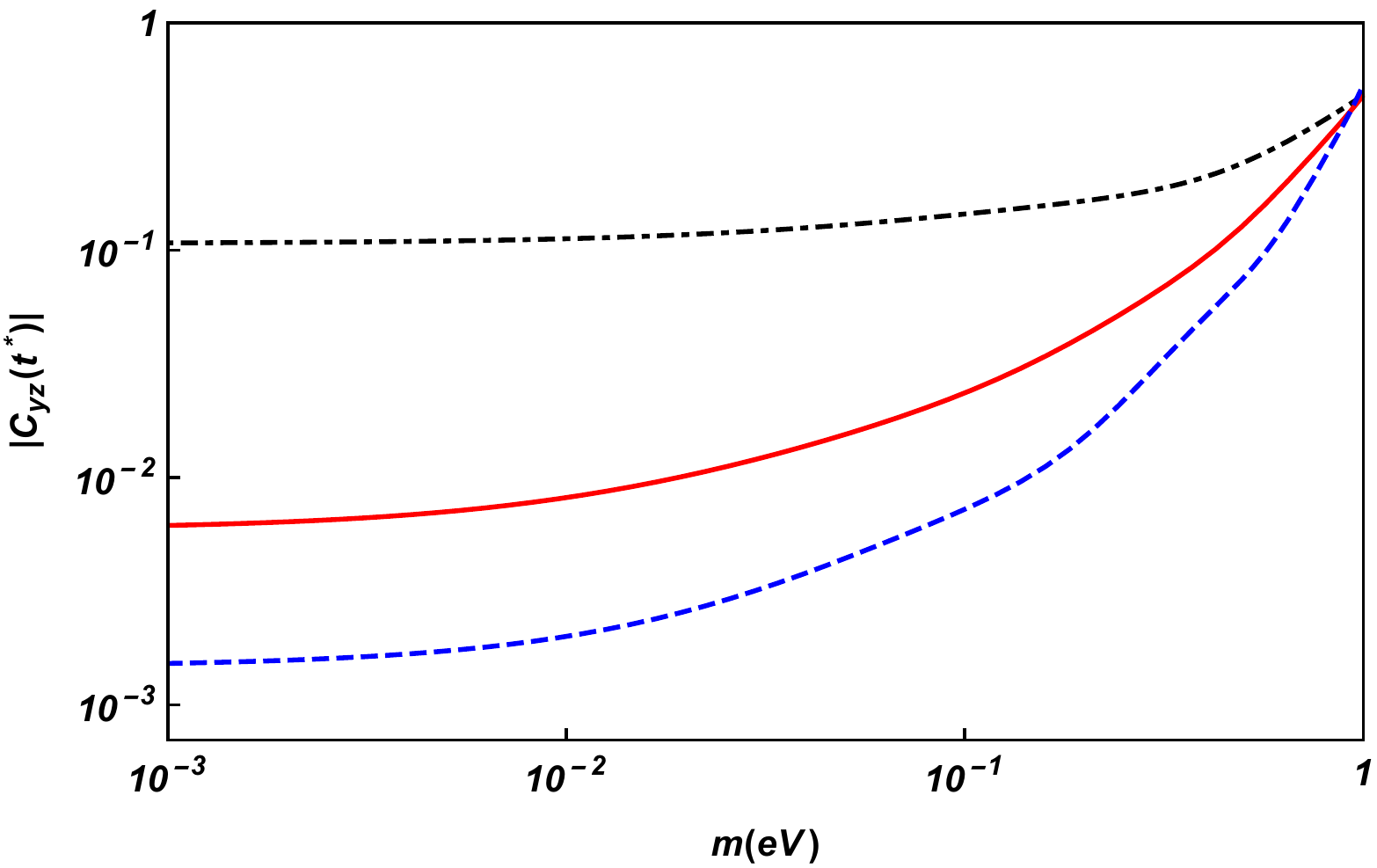}
\caption{
(color online) Plot of the $2$--Renyi entropies (upper panel) and of the Witnesses $C_{yz}$ (lower panel) at $t=t^*$, i.e. for $n=1$, between two neutrons as function of the axions mass $m$ at distance $r=1\, nm$.
In the plots we have considered two neutrons at a distance $r=1\, nm$ apart, and initial state given by Eq. \eqref{initial_state} with $\theta = \pi/4$ and $\phi = 0$. We set, for each value of the mass in the range $10^{-3} - 1$ eV, the coupling constant $g_p$ equal to the threshold value $g_{threshold}$ obtained from the experimental analyses in the refs. ~\cite{Bezerra2014,Bimonte2016,Chen2016,Klimchitskaya2015,Klimchitskaya2017}. In particular: for the black dot--dashed line we set $g_p = g_{CP}$, where $g_{CP}$ is the threshold from effective Casimir pressure measurements ~\cite{Bezerra2014}, and sample values are $g_{CP} = 0.0327$ for $m = 10^{-3} eV$, $g_{CP} = 0.0348$ for $m = 0.05 eV$, $g_{CP} = 0.0674$ for $m = 1 eV$; for the red solid line we set $g_p = g_{CF}$, where $g_{CF}$ is the threshold from measurements of the difference of Casimir forces ~\cite{Klimchitskaya2017}, and sample values are $g_{CF}=0.007$ for $m = 10^{-3} eV$, $g_{CF} = 0.012$ for $m = 0.05 eV$, $g_{CF} = 0.066$ for $m = 1 eV$; for the blue dashed line we set $g_p = g_{IE}$, where $g_{IE}$ is the threshold from isoelectronic experiments~\cite{Klimchitskaya2015}, and sample values are $g_{IE} = 0.0036$ for $m = 10^{-3} eV$, $g_{IE}=0.006$ for $m = 0.05 eV$, $g_{IE}=0.07$ for $m = 1 eV$.
}
\label{Graficinuovi}
\end{figure}

 For a numerical analysis of the $2$-Renyi entropy \eqref{2Renyi_entropy_tstar}, we focus on ALPs in the mass range $10^{-3} - 1$ eV. These have been considered in the refs. ~\cite{Bezerra2014,Bimonte2016,Chen2016,Klimchitskaya2015,Klimchitskaya2017}, where experimental constraints have been obtained on the coupling constant $g_{p}$ as a function of the ALP mass, both for protons and neutrons. In Fig.~\ref{Graficinuovi} we consider two neutrons at a distance $r=1\, nm$ apart, and initial state given by Eq. \eqref{initial_state} with $\theta = \pi/4$ and $\phi = 0$. We set, for each value of the mass in the range $10^{-3} - 1$ eV, the coupling constant $g_p$ equal to the threshold value $g_{threshold}$ obtained from the experimental analyses in the refs. ~\cite{Bezerra2014,Bimonte2016,Chen2016,Klimchitskaya2015,Klimchitskaya2017}. In particular: for the black dot--dashed line, we set $g_p = g_{CP}$, where $g_{CP}$ is the threshold from effective Casimir pressure measurements ~\cite{Bezerra2014}, and sample values are $g_{CP} = 0.0327$ for $m = 10^{-3} eV$, $g_{CP} = 0.0348$ for $m = 0.05 eV$, $g_{CP} = 0.0674$ for $m = 1 eV$; for the red solid line we set $g_p = g_{CF}$, where $g_{CF}$ is the threshold from measurements of the difference of Casimir forces ~\cite{Klimchitskaya2017}, and sample values are $g_{CF}=0.007$ for $m = 10^{-3} eV$, $g_{CF} = 0.012$ for $m = 0.05 eV$, $g_{CF} = 0.066$ for $m = 1 eV$; for the blue dashed line we set $g_p = g_{IE}$, where $g_{IE}$ is the threshold from isoelectronic experiments~\cite{Klimchitskaya2015}, and sample values are $g_{IE} = 0.0036$ for $m = 10^{-3} eV$, $g_{IE}=0.006$ for $m = 0.05 eV$, $g_{IE}=0.07$ for $m = 1 eV$. For the three cases, the $2$-Renyi entropy $S_2 (t^*)$ at $t=t^*$ is shown in the upper panel of Fig. ~\ref{Graficinuovi}.
All the lines depicted in the figure show a very similar behavior. At constant distance, larger axion masses imply a stronger damping by the Yukawa factor $e^{- mr}$.
Increasing the mass one eventually arrives at a value for which the Yukawa damping starts to be relevant.
From this point on, a further increment of the mass of the axions will suppress the entanglement exponentially.

To our knowledge, for ALPs in the mass range $10^{-3}-1 eV$, the values reported in refs.~\cite{Bezerra2014,Bimonte2016,Chen2016,Klimchitskaya2015,Klimchitskaya2017} represent the strongest model--independent constraints on axion--nucleon interactions from laboratory experiments. For this class of ALPs, as the figure~\ref{Graficinuovi} shows, the $2$-Renyi entropy is significantly different from zero, so that the laboratory constraints might be strengthened by several orders of magnitude from entanglement measurements. Of course, the so obtained constraints would be model--independent, since no specific axion or ALP model has been assumed. On the other hand, for QCD axion models there exist several constraints  from astrophysical sources, primarily from supernovae~\cite{Chang2018}, neutron star cooling~\cite{Hamaguchi2018} and Black Hole superradiance~\cite{Arvanitaki2015}. The indicative bound set by supernovae $g_{aNN}^2 \sim 10^{-19}$ renders our method unviable for QCD axions, at least for present day technologies. For general ALPs, since mass and coupling constants are essentially unrelated, and the latter can in principle assume any value, our approach can strengthen the current laboratory constraints, with the only limitations presented by the experimental sensitivities and coherence time.
It is worth to note that the   idea, here presented, to use entanglement to test theories of fundamental physics   is  in line with  several recent works, see for example Ref.~\cite{Marletto2017,Bose2017,Hsu2016} that suggest exploiting the entanglement as a probe for the quantum nature of the gravity.
In fact, in these papers, the spatial wave function has a non trivial evolution, and consequently, gravitational interactions can induce entanglement.

\section{The entanglement witness}

In the upper panel of Fig.~\eqref{Graficinuovi},
 we plot the  entanglement entropy between two neutrons as a function of the axions mass, at distance $r = 1nm$ and for
ALPs with coupling constants $g_p$ constrained by the analysis of Refs.~\cite{Bezerra2014},~\cite{Klimchitskaya2017} and~\cite{Klimchitskaya2015}.
Notice that a direct measurement of the 2--Renyi entropy is in general not easy to accomplish.
To overcome this  difficulty we can make use of an entanglement witness,
which is a quantity strictly related to the family of states and to the dynamics of the system under analysis,
whose value is able to signal the presence of entanglement.
Let us set to zero the phase $\phi$ of the initial state in eq.~(\ref{initial_state}). In this case, the entanglement witness can be identified with the two--spins correlation functions
 \begin{eqnarray}
 C_{yz}=\bra{ \psi(t)} \sigma_1^y \sigma_2^z \ket{\psi(t)} \equiv \bra{ \psi(t)} \sigma_1^z  \sigma_2^y \ket{ \psi(t)}.\label{witness_2}
 \end{eqnarray}
For any choice of $\phi$, one can find a different correlation function playing the same role.
It is straightforward to show that, for any time greater than zero, the  correlation function $C_{yz}$ is given by \mbox{$C_{yz}(t) =-\sin(2\theta) \sin\left(\Gamma t\right)$. This function, for any $\theta \neq \frac{k \pi}{2}$} with $k$ integer, vanishes only when the entanglement is zero.
For $t=nt^*=\frac{n \pi r^3}{6 A}$, the correlation function $C_{yz}$ depends only on the interaction between axions and fermions and reduces to
\begin{eqnarray}
\!C_{yz}(nt^*) \!\!&\!=\!&\!\!-\!\sin(2\theta)\! \sin\!\left[\!n\pi\! \left(\! 1  + \frac{B}{3} e^{-\! mr} (3\! + \!3mr\! + \!m^2r^2) \!\right)\! \right] \;   \nonumber  \\
 \!\!\!&\!\simeq \!&\!\!(-1)^{n  + 1} \! \sin(2\theta) n\pi B e^{-rm} \!\left(1\!+\!rm \!+\!\frac{r^2m^2}{3}\right) \,. \label{witness_3}
\end{eqnarray}
Therefore, the detection of such a quantity could demonstrate the existence of axions.

From eq.~(\ref{witness_3}), we can see that the entanglement witness, for $n \pi B\ll 1$, has a dependence on the interaction strength proportional to $g_p^2$ rather than to $g_p^4$ making the signal larger (indeed $B= \frac{4 g_p^2 }{\alpha g^2}$). In fact, the entanglement witness of our system assumes values larger than the entropy, making the detection of the axion and of the ALPs much more viable. This is shown in the lower panel of Fig.~\eqref{Graficinuovi}, where plots of the entanglement witness for $t= t^*$ are reported, and compared with those of the $2$-Renyi entropy in correspondence with the same parameters.

\section{Discussions and Conclusions}

In our analysis on the dynamics induced, in a system of two spin-$\frac{1}{2}$ fermions, by the axion--mediated fermion--fermion interaction, we have taken into account two important  constraints: the finiteness of the coherence time ~\cite{Buchleitner} and the limitation coming from the Yukawa damping factor $e^{-m r}$, which strongly suppresses the axion--mediated interaction, outside a limited spatial region $r < \frac{1}{m}$.

Relatively to the first constraint, it is well known that one of the main problems, when we are interested in the experimental analysis of the entanglement of a system, is the finiteness of the coherence time.
In realistic systems, the coherent superposition characterizing the pure quantum states is destroyed by interactions that, unavoidably, every quantum system shares with the surrounding world.
In our computations, we have neglected such kind of interaction and we assumed that our system is completely isolated from the rest of the universe.
This assumption is realistic since we have considered a finite time interval which is lesser than the coherence time, which in modern experimental setups can reach values of order of the second~\cite{Abobeih2018} and is continuously extending due to the progress in the experiments.
The characteristic time interval of our system, which has to be compared with the coherence time, is the minimum time interval needed to isolate the axion contribution to the entanglement ($t^{*}$).
For a system of two electrons a value of $t^*=7\,s$, which is at the limit of today's technology, is obtained by considering the relative distance $r=0.1 \, \mu m$.
A similar result is achieved for a system of two neutrons by considering relative distances of order of the nanometer.
On the other hand, the Yukawa damping factor has the only effect of reducing the range of distances $r$ for which the entanglement is significantly different from zero. This is not a serious limitation, since the smaller the axion mass, the larger is the spatial region  (with  $r < \frac{1}{m}$) where the model is efficient.

The knowledge of $t^*$ with a high precision is crucial for the approach described in the paper.
The largest source of error that can affect this quantity is represented by the uncertainty $\Delta r$ on the distance between the two fermions.
A simple analysis based on the definition of the entanglement witness in eq.~\eqref{witness_2} shows that an uncertainty $\Delta t^*$ on $t^*$ implies an uncertainty $\Delta C_{yz}(t^*)$ on the entanglement that, to lowest order in $B$, is equal to $\Delta C_{yz}(t^*) = \frac{\Delta t^*}{t^*} \sin(2\theta)$. The higher order terms $\mathcal{O}(B^2)$ can be safely neglected in virtue of the smallness of the coupling constant $g_p$. Any error $\Delta r$ on the inter--fermion distance then leaads to an uncertainty $\Delta C_{yz}(t^*) = 3\frac{\Delta r}{r} \sin(2\theta)$. From the definition of $t^*$ we can see that another source of uncertainty is represented by the fermion magnetic moments, which are known only with finite precision. Nevertheless, we expect that the uncertainty on the distance $\frac{\Delta r}{r}$ dominates over the uncertainty on the magnetic moments, which, in comparison, are known to a high degree of precision~\cite{Tanabashi2019}.
Given these considerations, a promising framework for the realization of our approach are optical lattices~\cite{Bloch2005,McGuyer2015}. In this context it is indeed possible to directly access spin correlation functions~\cite{Parsons2016}, with single site resolution imaging of fermions~\cite{Haller2015} and a precise control over the particle spacing.

In conclusion, taking into account the above constraints, we have analyzed the dynamics induced, in a system of two spin-$\frac{1}{2}$ fermions, by the axion--mediated fermion--fermion interaction in the non--relativistic regime.
We have shown that it is characterized by the rising of entanglement between the two fermions.
Moreover, we have shown that, by suitably tuning the observation time $t=nt^*$ and the distance between the two fermions, one can get rid of the contribution given to entanglement by the dipole--dipole interaction of magnetic origin.
In this way, any residual entanglement--entropy can be seen as a direct consequence of the presence of axions and hence constitutes a proof of their existence.
On the other hand if such entanglement is not detected this observation can be used to strengthen the current constraints.

In addition, to overcome possible difficulties in direct entropy measurement,
we have introduced a spin--spin correlation function which we have proved to be a suitable entanglement witness, that vanishes if and only if the entanglement goes to zero.
Such witness has also the advantage to be proportional, for $t=t^*$, to $g_p^2$ rather than to $g_p^4$ as the 2--Renyi entropy.
This fact allows extending the range of applicability of our experiment of several orders of magnitude.
The method we propose can likely probe a coupling constant range $g_p = 10^{-3}-10^{-1}$ for any axion mass up to $m \simeq 1 \, eV$, being particularly efficient for ALPs with low masses and large coupling constants~\cite{Marques2018}.
For coupling constants below $10^{-3}$ and masses beyond $m \simeq 1 \, eV$ measurements are limited by the current experimental precision.
Improvements in this respect may render wider regions of parameter space accessible in the next future.

\section*{Acknowledgments}
--
A.C.  G.L. and A.Q. thank partial financial support from MIUR and INFN. A.C. and G.L. thank also the COST Action CA1511 Cosmology and Astrophysics Network for Theoretical Advances and Training Actions (CANTATA).
SMG acknowledges   the European Regional Development Fund the Competitiveness and Cohesion Operational Programme (KK.01.1.1.06--RBI TWIN SIN), the Croatian Science Fund Projects No. IP-2016--6--3347 and IP-2019--4--3321, and
 the QuantiXLie Center of Excellence, a project co--financed by the Croatian Government and European Union through the European Regional Development Fund--the Competitiveness and Cohesion Operational Programme (Grant KK.01.1.1.01.0004).

\end{document}